# Pwning Level Bosses in MATLAB: Student Reactions to a Game-Inspired Computational Physics Course

Ian D. Beatty and Lauren A. Harris

*Department of Physics and Astronomy, University of North Carolina at Greensboro, PO Box 26170, Greensboro, NC 27402-6170*

**Abstract:** We investigated student reactions to two computational physics courses incorporating several videogame-like aspects. These included use of gaming terminology such as "levels," "weapons," and "bosses"; a game-style point system linked to course grades; a self-paced schedule with no deadlines; a mastery design in which only entirely correct attempts earn credit, but students can retry until they succeed; immediate feedback via self-test code; an assignment progression from "minions" (small, focused tasks) to "level bosses" (integrative tasks); and believable, authentic assignment scenarios. Through semi-structured interviews and course evaluations, we found that a majority of students considered the courses effective and the game-like aspects beneficial. In particular, many claimed that the point system increased their motivation; the self-paced nature caused them to reflect on their self-discipline; the possibility and necessity of repeating assignments until perfect aided learning; and the authentic tasks helped them envision using course skills in their professional futures.

## INTRODUCTION & THEORY

We report on a pilot study investigating whether and how incorporating video-game learning dynamics into face-to-face computational physics courses impacts student engagement and learning.

Many scholars have recognized that good video games are in fact highly effective learning environments, whether or not the content they teach is useful outside the game [1-6]. This realization has triggered an explosion of research into games designed to teach academic content [7-11], as well as experiments to make classroom-based courses more "game-like" [12,13]. Unfortunately, many of the latter merely appropriate superficial surface features of games. No consensus exists on what characteristics make a learning environment deeply "game-like."

In response, one of us (Beatty) has developed a theoretical model of "games as learning systems" to inform game-based and game-inspired instructional design [14-16]. This rather complex model identifies the key elements and dynamics that enable good video games to function as effective learning environments. At the heart of the model are three interwoven dynamics: an *intrinsic motivation loop* in which the player confronts an increasingly difficult sequence of challenges, motivated by feelings of achievement and growing confidence; an *exploratory learning loop* in which the player masters the knowledge and skills to overcome each particular challenge through goal-directed trial-and-error experimentation; and an *identity growth loop* in which the player experiences new ways of seeing and being in the world. The model suggests structural features that a game should have in order to support these dynamics, such as copious and timely positive feedback to promote feelings of achievement; minimal-consequence failure and immediate, informative feedback to support exploratory learning; and story details that connect the player to her in-game identity.

Beatty regularly teaches UNCG's two Computational Physics courses. During the 2013-2014 academic year, he taught redesigned versions incorporating several features intended to promote these loops. Our other author (Harris), an undergraduate teaching assistant for the course, studied students' thoughts about these features and reactions to the courses. This paper reports our preliminary results.

## CONTEXT

*Introduction to Computational Physics* (CP1) and *Computational Physics II* (CP2) are one-credit courses offered approximately biennially at UNCG. CP1 introduces students to *MATLAB*, the basics of procedural programming, a selection of numerical skills including statistical and graphical analysis, and random walk models. CP2 teaches numerical methods for analytically intractable physics problems, including integration of ordinary differential equations and mesh relaxation methods, as well as strategies for more complex programs with multiple interconnected pieces.

Beatty redesigned CP1 for the Fall 2013 semester and CP2 for the Spring 2014 semester to include six basic structural features that his model suggests will collectively and synergistically produce a trial-and-error exploratory learning loop coupled to an achievement-powered intrinsic motivation loop, while making a connection to students' nascent identities as computational physicists. He also included one cosmetic feature, the use of common video game terminology, to cue students to the other game-like characteristics of the course. This terminology included "levels" for course segments; "enemies," "minions," and "bosses" for assignments; "weapons" for useful *MATLAB* library functions and computational techniques; and "vanquishing" and "pwning" for successfully completing an assignment.

The first structural feature was the division of course material into "levels." Each consisted of nine to fourteen relatively short assignments called "minions" that introduced and developed specific *MATLAB* features, programming concepts, and numerical techniques, and one longer "level boss" that required integrating these to solve a more complex, less clear-cut problem. CP1 contained seven levels, while CP2 contained two of much greater scope and difficulty.

The second structural feature, intended to promote identity development, was framing level bosses as authentic computational physics tasks with a hypothetical setting and back-story that gave the student a role such as "summer intern at NASA."

The third structural feature was the expectation that students resubmit assignment solutions as often as necessary until succeeding, with only entirely correct solutions earning credit. (Grading of level bosses did include partial credit, but very little was earned for submissions that were not substantially correct.)

The fourth structural feature was the use of a game-style point system linked to course grades. In CP1, most minions were worth 10 points, with a few of unusual challenge ("lieutenants") worth 20. Level bosses were worth 100. Each level had a 20-point "speed bonus" earned by vanquishing all minions and obtaining at least 60% of the boss credit by a specified date; this was intended to help students pace themselves appropriately. In CP2, minions were worth 20 or 40 points, with bosses worth 200. CP2 included no speed bonuses, but did offer a 13-day "spring break special" with bonus points for assignments completed.

Every student began the course as a "rank 1" computational physicist, and "leveled up" one rank for every 80 points (CP1) or 40 points (CP2). A student's final course grade depended entirely upon his or her rank achieved at the end of the term. For CP1, a rank of 8 earned a D–, and each successive rank earned one higher grade-step (D, D+, etc.) up to a maximum of A+. CP2 used the same system, but with D– at rank 7. The course management system's gradebook was configured to keep students appraised of their point total, current rank, and current grade at all times. Thus, a student could see that she was, for example, ten points shy of "leveling up" to rank 14 and earning at least a B– in the course. A student's course grade could only increase and never decrease.

The fifth structural feature was the availability of immediate, on-demand, diagnostic feedback through automatic self-test code. Each level included a package of testing routines that would, when executed, put the student's programs through a suite of automated tests. If a student's programs passed all tests, the code printed a message such as "All 12 minions were vanquished!" along with an ASCII character-art smiley-face. If they didn't, the resulting message indicated "You slew 10 minions, but 2 survived," accompanied by extensive diagnostic information. Students could run the tests as often as they wished during development, and were sternly discouraged from submitting a solution (but not from seeking help) until it passed all tests. Once they submitted a solution, the instructor would inspect the code itself for any untested errors or flaws before awarding credit.

The sixth structural feature was the absence of any deadlines other than the semester's end, making the courses entirely self-paced. Any assignment could be submitted right up until the end, and many students completed significant course work (and significantly raised their final grades) during finals week.

## DATA COLLECTION & ANALYSIS

At the end of each course, Harris invited all enrolled students to participate in an interview soliciting their thoughts and reactions. Students were promised that their responses would be anonymized to protect confidentiality, and shared with nobody even in anonymized form until after course grades were submitted. Volunteers were offered no compensation of any form. 11 out of 15 students in CP1 and 7 out of 13 in CP2 agreed and were interviewed. (We attribute the low level of acceptance, especially for CP2, to the fact that many students left town for the holiday or summer before we could secure an interview.)

The interview protocol for each course (too long to include here) was semi-structured, and designed to solicit students' overall reactions to the course design as well as their thoughts about each of the specific game-like features. The interviewer asked neutral follow-up questions such as "tell me more" to draw out additional explanations or clarifications as seemed appropriate. Interviews typically lasted 10-20 minutes. The interviewer later transcribed the audio-recording of each interview, and checked the transcription.

Transcripts were identified by code numbers and scrubbed of any identifying personal information.

Harris coded the CP1 transcripts using both pre-determined and emergent codes. Pre-determined codes represented the game-like design features described above. Emergent codes were defined in response to common themes and significant ideas in participants' responses, and were intended to capture the range of their reactions with minimal analytical bias. The code set was refined through a process of "constant comparison" [17] in which transcripts were coded in succession, and then re-coded multiple times while tentative emergent codes were postulated, evolved, and refactored until they stabilized on a parsimonious set that adequately fit the apparent content. We checked inter-rater reliability by having a second analyst, unconnected with the course or students, independently code a subset of the transcripts. Five codes with particularly low values of Cohen's kappa were eliminated: four emergent codes, and one pre-determined ("general game-like nature"). The fourteen remaining codes had kappa values ranging from 1 (perfect) to 0.33 (fair), averaging 0.67 (substantial).

The final code set contained seven codes referring to course design features and seven emergent codes referring to students' various experiences, reactions, opinions, and observations. The design feature codes can be summarized as: game-like terminology; levels and assignment progression; complex authentic level bosses; infinite retries; point system; rapid feedback from self-test code; and self-paced. The emergent codes can be summarized as: addictiveness (e.g., not wanting to leave when class ends or staying awake late to level up); development of general programming ability; peer interaction (camaraderie and peer support); increased content learning ("more" than in a "normal" course); increased motivation; preparation for real-world work; and self-discipline (improving, or recognizing a lack).

Analysis of CP2 interview data is ongoing and will not be discussed in this paper. A second data source for both courses is the anonymous course evaluation questionnaire completed by most students at the end of each course. The questionnaire is generic and not tuned to these courses or this study, but does provide valuable corroboration.

## RESULTS

The first question in the CP1 interview protocol was "What is your overall opinion of the course?" Nine of eleven respondents answered with a generally positive statement, such as "I really love it. It's the only class that I have that like I'm sad when class is over and I just want to keep working on it," and "It's good. It's okay. I mean it's a programming course and it's self-paced, and I like that." One expressed a negative reaction ("For me, it required more effort than a three credit-hour course. Because I don't like computers. I think that's my personal situation"), and one an ambiguous reaction ("It's challenging. Um, I think that's about the only way I can describe it").

These responses suggest that most students appreciated the course design, an interpretation corroborated by results from the standard course evaluation questionnaire. In response to the prompt "What is your overall rating of this course?", eight of thirteen respondents rated the course 5/5 ("One of the best"), three rated it 4/5 ("Better than average"), and two rated it 3/5 ("About average"). For the prompt "What is your overall rating of this instructor's teaching?", eleven chose 5/5 ("Almost always effective"), and two chose 3/5 ("Sometimes effective"). Since the instructor did very little direct instruction, these responses must refer to some combination of the course design, assignment design, and individual assistance. For CP2, the equivalent counts were {6,3,1} and {7,2,1} out of 10 responses.

The second interview question was "Do you find the course's design increases or decreases your motivation?" Eight of the eleven interviewees responded affirmatively, ranging from "It doesn't make me hate the course, and believe me, I've had courses that have made me hate the course," to "Absolutely increases my motivation… I tend to do more for my computational class than for my other ones because I want the next rank." The other three responses were neutral ("I don't think it really affects my motivation") or mixed ("I enjoy the subject but it's very easy to put aside so I am less motivated to do it outside of class than my other work… In class it's great").

Again, the course evaluation results provide corroboration. For CP1, eleven of thirteen students responded to the open-ended prompt "What do you like most about this course and/or the instructor's teaching of it?" Of those, ten explicitly referenced game-like features of the course design or the instructor's attempt "to make the course fun." For CP2, all ten respondents wrote a response to that prompt; seven cited game-like aspects of the course or its general game-like nature, and two others identified the "larger", "real-world" level boss tasks.

Anecdotal observations recorded in Harris' field notes also reveal students' general positive affect during the courses. Students were often seen laughing or pumping a fist in the air when their program passed the self-test code and earned a smiley-face. In the final class meeting of CP2, students surprised the instructor with a set of matching, custom-made T-shirts bearing "All enemies were vanquished!" and the ASCII character-art smiley-face from the test code. Students would often comment on the video-game nature of the course, for example by saying "If it's a real game it should have cheats," and "I know why this class goes by so fast: It's a video game!"

Transcript locations where theory-driven codes (referring to course design features) co-occur with emergent codes (referring to student reactions) suggest that students perceive a causal relationship between the design feature and the reaction. A close reading of these transcript locations allows us to verify that the speaker is in fact intending such a causal claim. Four feature/reaction code pairs each co-occurred in at least five different participants' interview responses.

The most prevalent co-occurrence was the conjunction of "self-paced" with "self-discipline," explicitly mentioned by seven of the eleven respondents. An example quote so coded is "Yeah, it [the freedom to go at your own pace] means the student needs to take responsibility for their own work and I guess that's a good thing. I don't know if we're ready for it but we're going to have to in the industries out there so it's probably good. Even though I hate to admit it, yeah." Another example is "[A drawback of the self-paced aspect is that] it seems that all my other work that actually has due dates takes priority."

"Infinite retries" and "increased content learning" co-occurred in six interviews. Example quotes are "I find it [infinite retries] to be helpful because it forces you to actually learn what is going on," and "You actually have to learn it. You have to learn all the stuff." Some respondents contrasted CP1 with typical courses, in which "You're just supposed to never learn the stuff that you were supposed to learn in the first place. You just never, I mean, you miss the points and that's it forever."

A third common co-occurrence was "point system" with "increased motivation." Five respondents explicitly mentioned the motivating effect of the points-and-rank system, with two mentioning it for three different interview prompts. Typical quotes are "I will be at home and doing homework for other classes but all I'm thinking about is the fact that I only need another 30 points to rank up again in computational," and "It motivates me to get to that next level, to do the next thing and to increase my grade."

Equally common was the co-occurrence of "complex authentic level bosses" with "preparation for real-world work," also occurring in five interviews. An exemplar is "I like the real-life aspect of it just because it's uh an application, it's a real-life application, it's something you might see out in the real world."

## CONCLUSIONS & DISCUSSION

This is a preliminary report. Coding of the CP2 interviews and further analysis are in progress. Our sample of students is small, so statistical measures such as inter-rater reliability kappa values are suspect, and our analysis is predominantly based on self-reporting. Therefore, our findings may not generalize well to other populations; any specific result (e.g., the fact that many students believe that the "infinite retries" policy leads to increased content learning) should be considered suggestive only; and the causal relationships that students asserted might be more perceived than real.

Nevertheless, taken as a whole, the data and analysis results make a strong case that the game-like design of CP1 and CP2 was quite successful, at least for this particular population. Most students liked it, some fanatically; a few were indifferent; and none actively disliked it. Furthermore, most could articulate specific ways in which they thought specific game-like features of the design benefitted them.

One important question for future research is whether a similarly game-inspired course design can be equally successful for less skill-based, more concept- and formalism-focused courses such as introductory physics or upper-level theory.